\def\IEEEversion{0}
\if\IEEEversion1
    \documentclass[conference,letterpaper]{IEEEtran}
    \usepackage[cmex10]{amsmath}
\else
    \documentclass[english]{smfart}
    \usepackage{fullpage}
\fi
 
\usepackage[utf8]{inputenc} 
\usepackage[T1]{fontenc}
\usepackage{url}
\usepackage{ifthen}
\usepackage{mathrsfs}
\usepackage{euscript}
\usepackage{amssymb}
\usepackage{mathtools}
\usepackage{bm}
\usepackage{tikz}
\usepackage{enumerate}
\usepackage{algorithm, algorithmic}
\usepackage{color}
\usepackage{multicol}
\usepackage{multirow}
\usepackage{graphicx}
\usepackage{placeins}
\usepackage{hyperref}

\usepackage{mathrsfs}

 \newtheorem{definition}{Definition}
 \newtheorem{theorem}[definition]{Theorem}
 \newtheorem{proposition}[definition]{Proposition}

 \newtheorem{remark}{Remark}

\renewcommand{\leq}{\leqslant}

\newcommand{\eqdef}{\stackrel{\text{def}}{=}}

\newcommand{\F}{\mathbb{F}}
\newcommand{\Fq}{\mathbb{F}_q}

\newcommand{\Fqn}{\mathbb{F}_{q^n}}

\newcommand{\code}[1]{\mathscr{#1}}
\newcommand{\CC}{\code{C}}
\newcommand{\CCv}{\code{C}^{\text{vec}}}
\newcommand{\CCm}{\code{C}^{\text{mat}}}
\newcommand{\DC}{\code{D}}

\newcommand{\map}[4]{\left\{
\begin{array}{ccc}
#1 & \longrightarrow & #2 \\
#3 & \longmapsto     & #4
        \end{array}
\right.}

\newcommand{\word}[1]{\ensuremath{\boldsymbol{#1}}}
\newcommand{\Am}{\word{A}}
\newcommand{\Bm}{\word{B}}
\newcommand{\Cm}{\word{C}}
\newcommand{\Dm}{\word{D}}
\newcommand{\Em}{\word{E}}

\newcommand{\Mm}{\word{M}}

\newcommand{\Pm}{\word{P}}
\newcommand{\Qm}{\word{Q}}

\newcommand{\Xm}{\word{X}}
\newcommand{\Ym}{\word{Y}}

\newcommand{\av}{\word{a}}

\newcommand{\cv}{\word{c}}

\newcommand{\gv}{\word{g}}

\newcommand{\xv}{\word{x}}

\DeclareMathOperator{\rank}{Rank}

\newcommand{\Mspace}[2]{\mathcal{M}_{#1}(\Fq)}
\newcommand{\Mn}{\Mspace{n}{\Fq}}
\newcommand{\GL}[2]{\mathbf{GL}_{#1}(#2)}
\newcommand{\symn}{\text{Sym}_n(\Fq)}
\newcommand{\spanFqn}[1]{{\left\langle #1 \right\rangle}_{\Fqn}}
\newcommand{\inlinespanFqn}[1]{{\langle #1 \rangle}_{\Fqn}}

\newcommand{\qp}{\mathcal{L}_0}
\newcommand{\qqp}{\mathcal{L}}
\newcommand{\symnu}{\text{Sym}_{u}(\qqp)}
\newcommand{\gab}[1]{\mathbf{Gab}_{#1}}

\newcommand{\Tr}{\text{Tr}}
\newcommand{\Nr}{\text{N}}

\newcommand{\pair}[2]{{\langle #1, #2 \rangle}}
\newcommand{\pairu}[3]{{\langle #1, #2 \rangle}_u}

\begin{document}

\title{Improved decoding of symmetric rank metric errors} 

\if\IEEEversion1
 \author{
   \IEEEauthorblockN{Alain Couvreur\IEEEauthorrefmark{1}}
   \IEEEauthorblockA{\IEEEauthorrefmark{1}
                     INRIA Saclay \& LIX CNRS UMR 7161, \\
                     1 rue Honoré d'Estienne d'Orves,\\
                     91120 Palaiseau Cedex
                      France,\\
                     \texttt{alain.couvreur@inria.fr}}
 }
 \else
 \author{Alain Couvreur}
 \address{Inria \& LIX CNRS UMR 7161, \'Ecole Polytechnique,
   1 rue Honoré d'Estienne d'Orves, 91120 Palaiseau Cedex}
 \email{alain.couvreur@inria.fr}
 \fi
 \maketitle

\begin{abstract}
  We consider the decoding of rank metric codes assuming the error
  matrix is symmetric. We prove two results. First, for rates $<1/2$
  there exists a broad family of rank metric codes for which any
  symmetric error pattern, even of maximal rank can be corrected.
  Moreover, the corresponding family of decodable codes includes
  Gabidulin codes of rate $<1/2$.  Second, for rates $>1/2$, we
  propose a decoder for Gabidulin codes correcting symmetric errors of
  rank up to $n-k$. The two mentioned decoders are deterministic and
  worst case.
\end{abstract}

 \section*{Introduction}
Introduced in \cite{D78} by Delsarte and revisited by Gabidulin in
\cite{G85} and independently by Roth in \cite{R91}, rank metric codes
became a crucial object in information theory, with applications to
space--time coding, network coding, distributed storage or public key
cryptography. See for instance \cite{BHLPRWZ22a} for further details.
In short, a rank--metric code is a space of matrices endowed with the
following metric: the distance between two matrices is the rank of
their difference. Most of the studied rank metric codes in the
literature are the so--called {\em $\Fqn$--linear codes}.  These codes
are defined as follows. From an $\Fqn$--linear subspace of vectors
$\CCv$ of $\Fqn^n$, we deduce a space of matrices $\CCm$ by
``unfolding'' vectors in $\Fqn^n$ into $n\times n$ matrices with
entries in $\Fq$. Focusing on $\Fqn$--linear codes is somehow
restrictive since arbitrary matrix codes are only $\Fq$--linear and do
not have any $\Fqn$--linear structure. However, the most famous
rank-metric codes: Gabidulin codes, are $\Fqn$--linear.  These codes
are usually presented as a rank metric counterpart of Reed--Solomon
codes and benefit from similar features: for instance they are Maximum
Rank Distance (MRD), {\em i.e.} they reach the Singleton bound but
also they benefit from an efficient decoding algorithm correcting up
to half the minimum distance \cite{L06a,ALR13,GP08}. On one hand, the
Gabidulin decoder has the very same flavour as the Reed--Solomon
one. On the other hand, the decoding algorithm for Reed--Solomon codes
extends naturally to Sudan and Guruswami--Sudan decoding algorithms
while for Gabidulin codes the $\frac{n-k}{2}$ barrier seems very
difficult to pass through.  In \cite{RW15}, Raviv and Wachter--Zeh
proved that it was hopeless to expect an analogue of Sudan algorithm
for Gabidulin codes. More precisely, they proved that there could not
exist a {\em worst case} list decoder correcting beyond half the
minimum distance. The latter result does not however exclude the
existence of a decoder correcting beyond half the minimum distance
that may fail on some specific instances. This problem remains widely
open.

\subsection*{Previous work}
In another line of works, some authors focused on this decoding
problem under some structure assumption on the error. In \cite{GP06},
Gabidulin and Pilipchuk considered the specific case where the error,
when regarded as a matrix, is symmetric. In this setting, they
asserted that correcting such symmetric errors of weight beyond the
error correcting capacity of the code is possible.  However, some of
their result rest on assumptions on the minimum distance of some
ancilla code for which no evidence is given. In addition, they do not
provide a decoding algorithm. More recently, Jerkovits, Sidorenko and
Wachter--Zeh \cite{JSW21} considered a more general setting where the
row and column spaces of the error are equal. They call such errors
{\em space symmetric errors}. In this setting they proved that a
Gabidulin code can correct symmetric error up to rank
$\frac{2}{3}(n-k)$ instead of $\frac{1}{2}(n-k)$ with the usual
decoder. Their decoding algorithm is related to the decoding of
interleaved codes and may fail for some instances.  Finally, in
another direction the decoding of Gabidulin codes up to half the
minimum distance plus one for some specific structured error
error has been investigated in \cite{K21}.

\subsection*{Contributions of the present article}
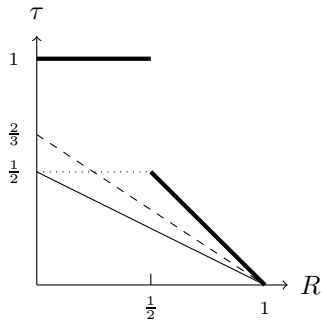
\begin{figure}[!h]
  \centering
  \begin{tikzpicture}[scale=1.5]
    \draw[->] (0,0) -- (2.2,0);
    \draw[->] (0,0) -- (0,2.2);
    \node at (0, 2.4){$\tau$};
    \node at (-.2, 2) {$\scriptstyle 1$};
    \node at (2.4, 0) {$R$};
    \node at (2, -.2) {$\scriptstyle 1$};
    \node at (1, -.2) {$\scriptstyle \frac{1}{2}$};
    \node at (-.2,1) {$\scriptstyle \frac{1}{2}$};
    \node at (-.2,1.333) {$\scriptstyle \frac{2}{3}$};
    \draw (1,0) -- (1,0.1);
    \draw[ultra thick] (0,2) -- (1,2);
\draw[ultra thick] (1,1) -- (2,0);
    \draw[dotted] (0,1) -- (1,1);
    \draw (0,1) -- (2,0);
    \draw[dashed] (0,1.33) -- (2,0);
  \end{tikzpicture}
  \caption{Comparison of decoding radii. The $x$--axis represents code
    rates and the $y$--axis the relative decoding radius
    $\tau = \frac t n$, where $t$ is the number of corrected errors.
    The thin plain line is the decoding radius of the usual decoder
    for Gabidulin codes. The dashed line is the decoding radius of
    \cite{JSW21}.  The dotted line is obtained by Gabidulin and
    Pilipchuk in \cite{GP06} for rates $< 1/2$.  The thick broken line
    is the radius achieved in the present paper for a symmetric
    error.}
    \label{fig:1}
\end{figure}

In this note, we provide two results.  First, we prove that for codes
of rate $<1/2$, it is possible to correct any symmetric error pattern
whatever its rank. This is possible with a worst case algorithm ({\em
  i.e.} with an algorithm that never fails), moreover, the algorithm
works on a very broad family of codes which includes Gabidulin ones.
Second, we prove that for rates $> 1/2$ it is still possible to
correct symmetric errors of rank up to $n-k$ with a
decoder on Gabidulin codes. The latter decoder is
deterministic and never fails when applied to valid
instances. For rate larger than $1/2$, we achieve the best decoding
radius expected by Gabidulin and Pilipchuk in their article.  A
comparison of the various decoding radii of \cite{GP06}, \cite{JSW21}
and the present paper is given in Figure~\ref{fig:1}.  Note that our
comparison with \cite{JSW21} is not completely fair since they lie in
a more general context: that of {space symmetric errors}.

\if\IEEEversion0 Note that the decoding radius we achieve is not a
continuous function of the rate which is a bit surprising. This is
discussed further in Section~\ref{sec:discussion}.
\fi

\if\IEEEversion0
\subsection*{Organisation of the article}
Section~\ref{sec:nota} gathers the basic notation used in the sequel.
In Section~\ref{sec:gabi} we recall some notions on $q$--polynomials
and Gabidulin codes and provide some material on the notion of
adjunction. Previous works are discussed in \ref{sec:previous}.
Finally, the contributions of the article are given in
Sections~\ref{sec:low} and~\ref{sec:high} where the decoding of
symmetric errors is studied first, for low rate codes and then for
high rate ones. Finally, the obtained decoding radius is discussed
in Section~\ref{sec:discussion}.
\fi

\section*{Acknowledgement}
The author is funded by the French {\em Agence nationale de la
  recherche} by the projects ANR-21-CE39-0009-BARRACUDA,
ANR-21-CE39-0011-COLA and {\em plan France 2030}, ANR-22-PETQ-0008.

 \section{Notation and prerequisites}\label{sec:nota}

\subsection{Matrix spaces}
The space of $n \times n$ matrices with entries in $\Fq$ is denoted as
$\Mn$.  A {\em code} is an $\Fq$--linear subspace of $\Mn$ endowed
with the rank metric, {\em i.e.} the distance given by
\[
  \forall \Am, \Bm \in \Mn,\ \ d(\Am, \Bm) \eqdef \rank (\Bm - \Am).
\]
For a given square matrix $\Mm$, its transpose is denoted by
$\Mm^\top$ and for a given matrix code
$\CC \subseteq \Mspace{n}{\Fq}$, we denote by $\CC^\top$ the {\em
  transposed code}, {\em i.e.} the space of transposed elements of
$\CC$. This code $\CC^\top$ has the same dimension as $\CC$ and should
not be confused with the dual code $\CC^\perp$ which is the orthogonal
space with respect to some non degenerate symmetric bilinear form.

Our focus is decoding in matrix spaces when the noise is
symmetric. More precisely, we wish to investigate the following
problem.  Given a code $\CCm \subseteq \Mn$, an integer
$0 \leq r \leq n$ and $\Ym = \Cm + \Em$ such that $\Cm \in \CCm$ and
$\Em$ is a {\bf symmetric} matrix of rank $r$, recover the pair
$(\Cm, \Em)$.

We will investigate this problem in two distinct contexts:
\begin{itemize}
\item matrix codes, {\em i.e.} $\Fq$--linear
  subspaces of $\Mn$
\item $\Fqn$--linear codes which are a specific case of matrix codes with
  an additional structure described below.
\end{itemize}

\subsection{$\Fqn$--linear vector codes}\label{ss:vec_codes}
$\Fqn$--linear codes are $\Fqn$--linear subspaces
$\CCv \subseteq \Fqn^n$. Choosing an $\Fq$--basis $\mathcal B$ of
$\Fqn$ yields a map $\Phi_{\mathcal B} : \Fqn^n\rightarrow \Mn$
sending $\xv = (x_1, \dots, x_n) \in \Fqn$ onto the matrix whose
$i$--th column are the coefficients of $x_i$ when decomposed in the
basis $\mathcal B$. This transform is not canonical since it depends
on the choice of the basis $\mathcal B$.
\if\IEEEversion0
However, suppose we consider
another basis $\mathcal B'$, then, for any $\cv \in \Fqn^n$,
$\Phi_{\mathcal B}(\cv) = \Pm \Phi_{\mathcal B'}(\cv)$ where
$\Pm \in \mathbf{GL}_n(\Fq)$ is the transition matrix from the basis
$\mathcal B$ to $\mathcal B'$. There remains to observe that the left
multiplication by $\Pm$ is a rank--preserving map and hence an
isometry with respect to the rank distance. Thus, a change of basis
yields the same matrix code up to isometry.
\else
However, the rank of $\Phi_{\mathcal B}(\cv)$ does not depend
on this choice.
\fi

\subsection{Code equivalence}
Two matrix codes $\CC, \DC \subseteq \Mspace{n}{\Fq}$ are said to be
{\em equivalent} if there exist $\Pm, \Qm \in \GL{n}{\Fq}$ such that
$\CC = \Pm \DC \Qm$, {\em i.e.} for any $\Dm \in \DC$, then
$\Pm \Dm \Qm \in \CC$ and conversely, any element of $\CC$ is of this
form. Equivalent codes are isometric with respect to the rank metric.
Finally, if $\Pm$ (resp. $\Qm)$ is the identity matrix, we say that the
codes are {\em right} (resp. {\em left}) {\em equivalent}.

 \section{$q$--polynomials and Gabidulin codes}\label{sec:gabi}

\subsection{$q$--polynomials}
Other crucial objects in the sequel are {\em $q$--polynomials}, which
provide a comfortable context to describe $\Fqn$--linear codes.  A
$q$--polynomial is a linear combination of $q^{\text{th}}$--powers of
the formal variable $X$. Namely it is a polynomial of the form
\begin{equation}\label{eq:qpoly}
  P(X) = a_0 X + a_1 X^q + \cdots + a_r X^{q^r}.  
\end{equation}
The integer $r$ is referred to as the {\em $q$--degree} of $P$ and denoted
$\deg_q(P)$. We endow $q$--polynomials with the addition and the
composition, yielding a noncommutative ring denoted by $\qp$.

\if\IEEEversion0
\begin{remark}
  The ring $\qp$ is usually known in computer algebra as the Ore ring
  \cite{O33} or the ring of twisted polynomials $\Fqn[T, \theta]$
  where $\theta$ denotes the Frobenius automorphism of $\Fqn$.
  Compared to usual rings of polynomials, in $\qp$ scalars do not
  commute with the variable and satisfy
  \[
    \forall a \in \Fqn, \ \forall i \in \mathbb N, \qquad
    T^i a = \theta^i (a) T^i = a^{q^i} T^i.
  \]
  Here we prefer notation \eqref{eq:qpoly} with $q$--th powers of $X$
  which is more usual in rank metric coding theory. 
\end{remark}
\fi

\subsubsection{$q$--polynomials as endomorphisms}
A $q$--polynomial na\-tu\-rally induces an $\Fq$--linear endomorphism of
$\Fqn$ and it is well--known that the quotient $\qqp \eqdef \qp/(X^{q^n} - X)$ of
$\qp$ by the two--sided ideal spanned by $X^{q^n}-X$ is isomorphic to
the whole space of $\Fq$--linear endomorphisms of $\Fqn$. Therefore,
after choosing an $\Fq$--basis of $\Fqn$, we deduce the existence of an
isomorphism
\begin{equation}\label{eq:iso_qpoly}
  \qqp \stackrel{\sim}{\longrightarrow} \Mspace{n}{\Fq}.
\end{equation}

For this reason, any matrix space can be regarded as an $\Fq$--linear
subspace of $\qqp$.

\begin{remark}
  A class $P \in \qqp$ as a unique representative of degree less than
  $n$.  In the sequel, we allow ourselves the abuse of language and
  notation to call {\em $q$--degree} of $P$ for the $q$--degree of
  this unique representative.
\end{remark}

\subsubsection{$q$--polynomials and $\Fqn$--linear vector codes}\label{sss:qpoly_codes}
An interesting feature of the algebra of $q$--polynomials and in
particular of the quotient algebra $\qqp$ is that it is $\Fqn$--linear
as a vector space. Hence it turns out to be a good setting to represent
$\Fqn$--linear vector codes as introduced in \S~\ref{ss:vec_codes}.

Indeed, given an $\Fq$--basis $\mathcal B = (g_1, \dots, g_n)$ of
$\Fqn$, we can establish an isomorphism
\[
\map{\qqp}{\Fqn^n}{P}{(P(g_1), \dots, P(g_n)).}
\]
The above map is well--defined. Let us explain why it is an
isomorphism.  If a $q$--polynomial $P$ vanishes at $g_1, \dots, g_n$,
then, by $\Fq$--linearity, it vanishes on the $\Fq$--span of the
$g_i$'s and hence on the whole $\Fqn$. From the isomorphism
(\ref{eq:iso_qpoly}), such a $P$ is zero in $\qqp$.
Finally, the equality of the $\Fq$--dimensions permits to deduce that
the map is an isomorphism.
More important, this map is $\Fqn$--linear and hence provides
an $\Fqn$--isomorphism between $\qqp$ and $\Fqn^n$.
Finally, one can easily check that the map is rank--preserving.
Thus, via such a map we establish a (non canonical) correspondence between
$\Fqn$--linear vector codes and $\Fqn$--linear subspaces of $\qqp$.

\if\IEEEversion0
\begin{remark}\label{rem:two-sides}
  Note that $\qqp$ is actually $\Fqn$--linear in two different
  manners.  It is $\Fqn$--linear on the left, the multiplication of
  $P\in \qqp$ by a scalar $\alpha \in \Fqn$ can be understood as the
  left composition by $\alpha X$, that is
  $\alpha \cdot P = \alpha X \circ P$.  On the other hand, one can
  also compose $P$ by $\alpha X$ {\em on the right}, which yields a
  distinct notion of $\Fqn$--linearity.  In the sequel, when
  discussing $\Fqn$--linearity and unless otherwise specified we
  always mean $\Fqn$--linearity {\em on the left}.
\end{remark}
\fi

Finally, for two $\Fqn$--linear codes
$\CC, \DC \subseteq \qqp$, the codes are {\em right equivalent} if
$\CC = \DC \circ P$ for some $P \in \qqp^{\times}$.

\subsubsection{Adjunction}
In the sequel, we wish to deal both with matrix codes and
$\Fqn$--linear codes which will be represented by $\Fqn$--linear
spaces of $q$--polynomials. When representing $q$--polynomials as
matrices, we would like to consider a relevant choice of basis for
which matrix transposition will correspond to a natural operation on
$q$--polynomials. For this sake we look for orthonormal bases with
respect to the trace bilinear form.

Recall that the trace and norm maps are defined
as
\if\IEEEversion1
\[
  \forall x \in \Fqn, \quad \Tr_{\Fqn/\Fq} (x) \eqdef x + x^q +
  x^{q^2}+\cdots + x^{q^{n-1}}
\]
and
\[
  \Nr_{\Fqn/\Fq}
  (x) \eqdef x x^q \cdots x^{q^{n-1}} = x^{\frac{q^n - 1}{q-1}}.
\]
\else
\[
  \forall x \in \Fqn, \quad \Tr_{\Fqn/\Fq} (x) \eqdef x + x^q +
  x^{q^2}+\cdots + x^{q^{n-1}} \quad \text{and} \quad \Nr_{\Fqn/\Fq}
  (x) \eqdef x x^q \cdots x^{q^{n-1}} = x^{\frac{q^n - 1}{q-1}}.
\]
\fi
Note that usually, the extension is specified as a subscript:
$\Tr_{\Fqn/\Fq}$ and $\Nr_{\Fqn/\Fq}$. However, since in this article,
the extension will always be $\Fqn/\Fq$ these
subscripts are omitted.

Next, we equip $\Fqn$ the following symmetric bilinear forms,
for any $u \in \Fqn^{\times}$
\begin{equation}\label{eq:traceform}
  \pairu{\cdot}{\cdot}{u} : \map{\Fqn \times \Fqn}{\Fq}{(x,y)}{\Tr(uxy).}
\end{equation}
When $u = 1$, we get the usual trace bilinear form which is denoted as
$\pair{\cdot}{\cdot}$.

It is proved in \cite{SL80} that when $q$ is a power of $2$ or when
both $q$ and $n$ are odd, then there exists an orthonormal basis for
the bilinear form $\pair{\cdot}{\cdot}$.  That is to say a basis
$x_1, \dots, x_n$ such that for any
$i,j,\ \pair{x_i}{x_j} = \delta_{ij}$, where $\delta_{ij}$ denotes the
Kronecker delta.  In the remaining case, {\em i.e.} when $q$ is odd
and $n$ is even, then replacing $\pair{\cdot}{\cdot}$ by
$\pairu{\cdot}{\cdot}{u}$ for a non square $u$, we get the existence of
an orthonormal basis as claimed in the following statement.

\begin{proposition}
  Suppose that the characteristic of $\Fq$ is different from $2$ and
  that $n$ is even. Then if $\Nr(u) \in \Fq^\times$ is not a square,
  the bilinear map $\pairu{\cdot}{\cdot}{u}$ has an orthonormal basis.
  \end{proposition}

  \if\IEEEversion1 \begin{IEEEproof} \else \begin{proof} \fi The
      bilinear form is non degenerate and by the classification of
      quadratic forms over finite fields, the existence of an
      orthonormal basis is equivalent to the fact that its
      discriminant is a square in $\Fq^\times$. Recall that the
      discriminant of a symmetric bilinear form is an element of
      $\Fq^{\times}/ {(\Fq^{\times})}^2$, hence it is defined up to a
      square.  Let $\av = (a_1, \dots, a_n) \in \Fqn^n$ be an
      $\Fq$--basis of $\Fqn$.  Following \cite[p.~62]{LN97}, it is
      known that the discriminant $\Delta$ of $\pair{\cdot}{\cdot}$
      can be obtained as the determinant of the matrix
      $\Am = \Mm(\av)^\top \Mm(\av)$ where $\Mm(\av)$ is the Moore
      matrix associated to $\av$:
    \[
      \Mm(\av) \eqdef
      \begin{pmatrix}
        a_1 & a_2 & \cdots & a_n\\
        a_1^q & a_2^{q} & \cdots & a_n^q \\
        \vdots & \vdots & & \vdots \\
        a_1^{q^{n-1}} & a_2^{q^{n-1}} & \cdots & a_n^{q^{n-1}}
      \end{pmatrix}.
    \]
    Therefore, $\Delta = \det(\Mm (\av))^2$ but, from \cite{SL80},
    $\det(\Mm(a))$ is actually in $\F_{q^2} \setminus \Fq$ and hence
    $\Delta$ is not a square in $\Fq$.

    Now, the discriminant $\Delta_u$ of $\pairu{\cdot}{\cdot}{u}$
    equals $\det (\Mm(\av)^\top \Dm_u \Mm(\av))$ where
    \[
      \Dm_u \eqdef
      \begin{pmatrix}
        u & & & (0)\\
        & u^q & & \\
        &   & \ddots & \\
        (0)&   &        & u^{q^{m-1}}
      \end{pmatrix}.
    \]
    Therefore, for any $u \in \Fqn$,
    $\Delta_u = \det(\Mm(\av))^2 \det(\Dm_u)$. There remains to
    observe that $\det(\Dm_u) = \Nr(u)$. Consequently, if $\Nr(u)$ is a
    non square, then $\Delta_u$ is a square.
    \if\IEEEversion1 \end{IEEEproof} \else \end{proof} \fi

\if\IEEEversion0
\begin{remark}
  Note that the norm map sends squares onto squares and non squares
  onto non squares. Therefore, the condition $\Nr(u)$ is a non square
  in $\Fq^\times$ is equivalent to the condition $u$ is a non square
  in $\Fqn^\times$.
\end{remark}
\fi

The bilinear form $\pairu{\cdot}{\cdot}{u}$ is known to be non
degenerated and with this pairing at hand one can define a notion of
adjunction for endomorphisms and hence for elements of $\qqp$.  
Namely, given $P \in \qqp$, we define
$P^{\top,u} \in \qqp$ the unique element of $\qqp$ satisfying
\[ \forall a, b \in \Fqn, \qquad
  \Tr(aP(b)) = \Tr(P^{\top,u}(a)b).
\]

One sees easily that for any $a \in \Fqn$, the multiplication by $a$
map, corresponding to the $q$--polynomial $aX$, is self--adjoint. Next,
for any $i \in \{1, \dots,m-1\}$, one can prove that
${(X^{q^i})}^{\top,u} = \frac{u^{q^{m-i}}}{u} X^{q^{m-i}}$. Finally, one
easily proves that for any $P, Q \in \qqp$, we have an
anticommutativity property, namely: $(PQ)^{\top,u} = Q^{\top,u}
P^{\top,u}$. Putting all together, we get an explicit formula for the
adjunction operator. For
$P = p_0 X + p_1 X^q + \cdots + p_{m-1}X^{q^{m-1}}$, we have
\[
  P^{\top,u}(X) \eqdef p_0 X + \sum_{i=1}^{m-1} p_i^{q^{m-i}} u^{q^{m-i}
    - 1} X^{q^{m-i}}.
\]

Finally, notice that, representing a $q$--polynomial $P$ in an
orthonormal basis for $\pairu{\cdot}{\cdot}{u}$ yields a matrix $\Mm$,
and $P^{\top,u}$ will be represented by the transposed matrix $\Mm^\top$.

This encourages to define the adjoint code of a given code.

\begin{definition}\label{def:adjoint_code}
  Let $\CC \subseteq \qqp$ be an $\Fqn$--linear code (on the left), then
  one can define $\CC^{\top,u}$ to be the image of $\CC$ by the map
  $P \mapsto P^{\top,u}$.
\end{definition}

\begin{remark}
  If $\CC$ is $\Fqn$--linear on the left, then $\CC^{\top,u}$ is $\Fqn$--linear
  on the right.
\end{remark}

\begin{remark}
  Actually, the objective is to benefit from an adjunction operation
  such that $q$--polynomials $aX$ for $a \in \Fqn$ are
  self-adjoint and adjoints of monomials are collinear to monomials.
  Such an adjunction is useful in the sequel since we will see that it
  sends Gabidulin codes onto Gabidulin codes (up to right
  equivalence). When $q, n$ are both odd, one needs to use
  $\pairu{\cdot}{\cdot}{u}$ instead of the trace bilinear map, which
  seems a bit cumbersome.  However, we did not succeed in finding a
  non degenerate symmetric bilinear map for which $q$--polynomials of
  the form $aX$ for $a\in \Fqn$ are self-adjoint and monomials
  $X^{q^i}$ are ``orthogonal'' {\em i.e.}  are inverse to their
  adjoint.
  \if\IEEEversion0
  This is exactly what happens with the trace bilinear form
  when it has an orthonormal basis (for instance $q$ odd and $n$
  even). For the general case, we leave it as an open question.
  \fi
\end{remark}

\subsection{Gabidulin codes}
Gabidulin codes are a peculiar class of $\Fqn$--linear codes
discovered by Delsarte in \cite{D78} and then revisited by Gabidulin
in \cite{G85}. These codes can be regarded as rank--metric analogues
of Reed--Solomon codes and benefit in particular from an efficient
decoding algorithm correcting up to half the minimum distance
\cite{L06a,ALR13,GP08}. Usually these codes are regarded as subspaces
of $\Fqn^n$ and defined as follows. Consider an integer
$0 \leq k \leq n$ and a sequence $\gv = (g_1, \dots, g_n)$ of
$\Fq$--linearly independent elements of $\Fqn$. The code
$\mathbf{Gab}_k(\gv)$ is defined as \if\IEEEversion1
\begin{align*}
  \mathbf{Gab}_k(\gv) \eqdef \{(P(g_1), &\dots, P(g_n)) ~|~\\
  & P \in \qqp,\
  \deg_q(P) < k\}.
\end{align*}
\else
\[
  \mathbf{Gab}_k(\gv) \eqdef \{(P(g_1), \dots, P(g_n)) ~|~ P \in \qqp,\
  \deg_q(P) < k\}.
\]
\fi Since Gabidulin codes are $\Fqn$--linear, then, according to
\S~\ref{sss:qpoly_codes}, these codes can be regarded as
$\Fqn$--linear subspaces of $\qqp$ and this is the definition we will
use in the sequel.

\begin{definition}
  Let $0 \leq k \leq n$ be an integer. The $k$--dimensional Gabidulin code
  $\gab{k}$ is defined as
  \[
    \gab{k} \eqdef \{P \in \qqp ~|~ \deg_q P < k\}.
  \]
\end{definition}

\if\IEEEversion0
\begin{remark}
  Following the discussion started in Remark~\ref{rem:two-sides}, a
  Gabidulin code is two--sided $\Fqn$--linear.
\end{remark}
\fi

We finish with a statement on adjunction and Gabidulin codes.

\begin{proposition}
  Let $\CC = \gab{k}$, then, for any $u \in \Fqn^\times$, the code
  $\CC^{\top,u}$
  \if\IEEEversion0
  $\eqdef \{P^{\top,u} ~|~ P \in \CC\}$
  \fi
  is right equivalent to
  itself.
\end{proposition}

\if\IEEEversion1 \begin{IEEEproof} \else \begin{proof} \fi $\gab k$ is
    two--sided $\Fqn$--linear. Then, its adjoint $\gab{k}^{u,*}$ is
    two--sided linear too. Moreover,
  \[
    \CC^{\top,u} = \spanFqn{X, X^{q^{n-1}}, \dots, X^{q^{n-k+1}}}.
  \]
  Then,
  \[
    \CC^{\top,u} \circ X^{q^{k-1}} = \spanFqn{X, X^q, \dots, X^{q^{k-1}}} = \CC. 
  \]
\if\IEEEversion1 \end{IEEEproof} \else \end{proof} \fi

 \section{Previous works}\label{sec:previous}
In this section, we discuss two previous works on the decoding of
Gabidulin codes when the error is a symmetric matrix.  We first recall
the main idea of the Gabidulin decoder presented for instance in
\cite{L06a,ALR13}. We describe it using the representation of
$\Fqn$--linear codes by $q$--polynomials.

Given $Y = C+E$, where $Y \in \qqp$, $C \in \gab{k}$, {\em i.e.}
$\deg_q (C) < k$, and $E \in \qqp$ such that
$\rank (E) \leq t$.  The
decoding consists in choosing $L \in \qqp$ with $\deg_q (L) \leq t$
whose kernel contains the image of $E$ so that $L \circ E = 0$.  Such
a $q$--polynomial is usually referred to as a {\em localiser} of the
error since, it permits to localize the image of the error
$q$--polynomial $E$.  To do it, we solve the following linear system:
\begin{equation}\label{eq:WB}
  L \circ Y = N
\end{equation}
for $L, N \in \qqp$ such that $\deg_q (L) \leq t$ and
$\deg_q (N) \leq k+t-1$.

It can be proved that if
$t \leq \left\lfloor \frac{n-k}{2} \right\rfloor$, then any nonzero
solution $(L, N)$ actually equals $(L, L \circ C)$ and $C$ can
be recovered by left Euclidean division by $L$.

\subsection{Gabidulin and Pilipchuk}
In \cite{GP06}, the authors make the following observation.  Suppose
we received $Y = C+E$ where $C \in \gab{k}$ and $E \in \qqp$ is
symmetric ({\em i.e.}  self--adjoint with respect to
$\pairu{\cdot}{\cdot}{u}$). Then,
\[
  Y^{\top,u} = C^{\top,u} + E.
\]
Therefore, one can access to the syndromes of $E$ with respect to both
$\gab{k}$ and $\gab{k}^{\top,u}$. Then, it is as if they were decoding
$\gab{k} \cap \gab{k}^{\top,u}$ and the authors then mention that,
depending on the minimum distance of this latter code, it may be
possible to correct between $\frac{n-k}{2}$ and $n-k$ errors.
More precisely,
\begin{itemize}
\item for rates $< 1/2$, they observed that
  $\gab{k} \cap \gab{k}^{\top,u}$ is nothing but a one-dimensional
  Gabidulin code for which one can correct errors of rank up to
  $\frac{n-1}{2}$.
\item For rates $> 1/2$ they did not succed in identifying the
  structure of $\gab{k} \cap \gab{k}^{\top,u}$. They hence suggested a
  possible decoding radius lying between $\frac{n-k}{2}$ and $n-k$
  depending on the minimum distance of
  $\gab{k} \cap \gab{k}^{\top,u}$ but did not provided
  any decoding procedure in this case.
\end{itemize}

In \S~\ref{sec:high} we revisit their approach and prove that decoding
up to $n-k$ errors can always be achieved.

\subsection{Jerkovits, Sidorenko and Wachter--Zeh}
In \cite{JSW21}, the authors consider a more general setting and only
assume that the row space and the column space of the error coincide.
Their approach can be described as follows. The condition on $E$
entails that any localiser $L$ satisfies both $L \circ E = 0$ and
$E \circ L = 0$. Thus, it is possible to double the number of
equations of the system (\ref{eq:WB}) and get a system:
\begin{align*}
  L\circ Y & = N_1 \\
  Y \circ L & = N_2
\end{align*}
When solving the system, we get a unique solution for almost any entry
as soon as the rank of the error is below $\frac{2}{3}(n-k)$.

\begin{remark}
  This approach is strongly related to the decoding of
  interleaved Gabidulin codes. See for instance \cite{WZ14}.
\end{remark}

 \section{Decoding symmetric errors in rate $< 1/2$}\label{sec:low}
As explained earlier, rank metric codes can be represented as spaces
of matrices, spaces of vectors with entries in $\Fqn$ or spaces of
$q$--polynomials. In this section, since there is no need for an
$\Fqn$--linear structure, we will deal with arbitrary spaces of matrices.
In the sequel, denote by $\symn$ the space of $n\times n$ symmetric
matrices.

\begin{theorem}
  Let $\CC \subseteq \Mspace{n}{\Fq}$ such that
  $\CC \cap \symn = \{0\}$, then there is a polynomial time decoding
  algorithm which returns $\Cm$ for any input $\Cm + \Em$ where
  $\Cm \in \CC$ and $\Em \in \symn$.
\end{theorem}

\if\IEEEversion1 \begin{IEEEproof} \else \begin{proof} \fi
  From the input $\Ym = \Cm + \Em$ for $\Cm \in \CC$ and $\Em \in \symn$,
  we apply the map
  \begin{equation}\label{eq:map_Phi}
    \Phi : \map{\Mspace{n}{\Fq}}{\Mspace{n}{\Fq}}{\Xm}{\Xm - \Xm^\top.}
  \end{equation}
  Note that $\ker \Phi = \symn$ and hence
  $\Phi(\Ym) = \Phi (\Cm) = \Cm - \Cm^\top$.  Next, since
  $\ker \Phi = \symn$ the restriction of $\Phi$ to $\CC$ is injective.
  Therefore, $\Phi(\Ym) = \Phi(\Cm)$ has a unique pre-image by $\Phi$
  in $\CC$ that can be uniquely recovered by solving a linear system.
  \if\IEEEversion1 \end{IEEEproof} \else \end{proof} In terms of
complexity, the aforementioned algorithm only involves linear algebra
over the space of $n \times n$ matrices and hence costs
$O(n^{2\omega})$ operations in $\Fq$ where $\omega \leq 3$ denotes the
complexity exponent of linear algebra operations.  \fi

Let us conclude this section by a few comments about the
aforementioned decoder.
\begin{itemize}
\item The decoder is deterministic and {\em worst case}, for any
  codeword $\Cm$ corrupted by any symmetric error pattern, the
  algorithm returns $\Cm$ and never fails.
\item This holds for {\em any} symmetric error pattern, even with
  errors of rank $n$.
\item The algorithm works for a broad family of codes whose
  $\Fq$--dimensions are below
  $n^2 - \dim_{\Fq} \symn = n^2 - {n+1 \choose 2}$.  Asymptotically,
  this yields a broad family of codes with rate $<1/2$ which can be
  decoded by this way under the assumption that the error is
  symmetric.
\end{itemize}

To conclude, note that for any $k < \frac n 2$, a right equivalent
code of $\gab{k}$  lies in this family. Indeed,
choose
$\CC = \gab{k} \circ X^{q} = \inlinespanFqn{X^q , \dots, X^{q^k}}$.
Then $\CC^{\top,u} = \inlinespanFqn{X^{q^{n-k}}, \dots, X^{q^{n-1}}}$.
In particular, $\CC \cap \CC^{\top,u} = \{0\}$ and hence no element of
$\CC$ is symmetric with respect to $\pairu{\cdot}{\cdot}{u}$.

 \section{Decoding symmetric errors in rate $> 1/2$}\label{sec:high}

In this last section, we show that using Gabidulin codes of
$\Fqn$--dimension $k$ it is possible to achieve a decoding radius of
$n-k$, that is to say, up to the minimum distance minus one.

Here we still assume that errors are symmetric matrices but we switch
to a $q$--polynomial description. Thus we choose a bilinear form
$\pairu{\cdot}{\cdot}{u}$ having an orthonormal basis, so that any
matrix in $\Mspace{n}{\Fq}$ can be represented by a unique element of
$\qqp$ and any symmetric matrix $\Mm$ is represented by a
$q$--polynomial $M$ such that $M^{\top,u} = M$. We denote the
space of self-adjoint $q$--polynomials as $\symnu$.

\begin{theorem}
  Let
  \[
  \CC \eqdef \gab{k} \circ X^q = \spanFqn{X^q, \dots, X^{q^k}}
\]
and suppose we are given $Y = C+E$ where $E$ is self--adjoint with
respect to $\pairu{\cdot}{\cdot}{u}$ and $\rank (E) \leq n-k$. Then
there is a deterministic polynomial time algorithm returning $C$.
\end{theorem}

\if\IEEEversion1
\begin{IEEEproof}
\else
\begin{proof}
\fi
The first step of the decoder is similar to the low rate case, we start
by applying the operator $\Phi$ previously introduced. When applied to
$q$--polynomials it is defined as
\[
  \Phi : \map{\qqp}{\qqp}{M}{M - M^{\top,u}}
\]
We compute $\Phi(Y) = C + C^{\top,u}$. Now the restriction
$\Phi_{|\CC}$ of $\Phi$ to $\CC$ is no longer injective. Indeed, the
kernel of $\Phi_{|\CC}$ equals $\CC\cap \symnu$ which
satisfies
\if\IEEEversion1
\begin{align*}
  \CC \cap \symnu \subseteq \spanFqn{X^{q^{m-k}}, \dots, X^{q^k}},
\end{align*}
the right--hand side being equal to $\gab{2k - n + 1} \circ X^{q^{n-k}}$.
\else
\[
  \CC \cap \symnu \subseteq \spanFqn{X^{q^{n-k}}, \dots, X^{q^k}}
  = \gab{2k - n + 1} \circ X^{q^{n-k}}.
\]
\fi Therefore, after computing $\Phi(Y) = \Phi(C)$, pick a random
preimage of $\Phi(Y)$ by $\Phi$ in $\CC$. This yields $C' = C - S$ for
some $S \in \CC \cap \symnu \subseteq \gab{2k-n-1} \circ X^{q^{n-k}}$.

Then, compute, $Y - C' = S + E$.  Since
$S \in \gab{2k-n+1} \circ X^{q^{n-k}}$, one can apply the usual
decoding of Gabidulin codes to $S+E$ in order to recover $E$.
The aforementioned decoder corrects up to $\frac{2n-2k}{2} = n-k$
errors.
\if\IEEEversion1
\end{IEEEproof}
\else
\end{proof}
\fi

 \if\IEEEversion0
\section{Discussion about the obtained decoding radii}\label{sec:discussion}

It is legitimate to ask why our decoding radius curve
in Figure~\ref{fig:1} is discontinuous. However, this is not very surprising.
First, because the decoders for $R<1/2$ and $R>1/2$ are of different nature.
The former is based on the trivial operation $\Phi$ of \eqref{eq:map_Phi}
and corrects any error pattern whatever its rank.
It works for Gabidulin codes but actually also for a much broader family:
the matrix codes which do not contain nonzero symmetric matrices.
The latter decoder concerns only Gabidulin codes, it starts with the
same basic operation $\Phi$ as the former but then applies the actual
Gabidulin decoder.

Somehow, the discontinuity we observe in the decoding radius was
already known but not especially observed. Indeed, consider the zero
code, which can be regarded as the zero--dimensional Gabidulin
code. For this code, correcting any error pattern is trivial. Thus, it
corrects any error pattern of rank up to $n$.  Next, consider the
one--dimensional Gabidulin code. This code can only correct errors up
to rank $\frac{n-1}{2}$. This yields a discontinuity. Somehow, the
decoders we identify in the present article, take advantage of the
structure of the error in order to ``shift to the righft'' the
discontinuity at zero we just mentioned in the case of arbitrary
errors.

 \fi
\section*{Conclusion}
We provided a broad family of rate $<1/2$ codes for which the decoding
of symmetric errors is obvious, can be performed whatever the rank of
the error and succeeds for any error pattern. The aforementioned
family of codes codes is nothing but the matrix spaces which do not
contain any nonzero symmetric matrix. This family of includes
Gabidulin codes represented in adequate bases.

For codes of rate $> 1/2$, we give a decoding algorithm for Gabidulin
codes that correct any symmetric error pattern of weight up to
$n-k$. This algorithm never fails. It confirms the expectations of
Gabidulin and Pilichuk \cite{GP06} who mentioned that this decoding
radius could be reachable without providing a decoding algorithm for
that.

\if\IEEEversion1
\bibliographystyle{IEEEtran}
\else
\bibliographystyle{alpha}
\fi
\newcommand{\etalchar}[1]{$^{#1}$}

\end{document}